\documentclass[preprint2]{aastex}

\newcommand{\masy}{\mbox{mas y$^{-1}$}}

\slugcomment{Submitted to the AJ}

\shorttitle{Fornax PM}
\shortauthors{M\'endez et al.}

\begin{document}

\title{First ground-based CCD proper motions for Fornax II: Final
  results}

\author{Ren\'e. A. M\'endez and Edgardo Costa}
\affil{Departamento de Astronom\'ia, Universidad de Chile, Casilla
  36-D, Santiago, Chile}
\email{rmendez, costa@das.uchile.cl}

\author{Carme Gallart}
\affil{Instituto de Astrof\'isica de Canarias, Tenerife 38200, Islas Canarias, Spain}
\email{carme@iac.es}

\author{Mario H. Pedreros}
\affil{Departamento de F\'isica, Universidad de Tarapac\'a, Casilla 7-D, Arica, 
Chile}
\email{mpedrero@uta.cl}

\author{Maximiliano Moyano}
\affil{Max-Planck-Institut f\"ur Astronomie, K\"onigstuhl 17, D-69117,
  Heidelberg, Germany}
\email{moyano@mpia-hd.mpg.de}

\and

\author{Martin Altmann}
\affil{Zentrum f\"ur Astronomie,
  M\"onchhofstr. 12-14, 69120 Heidelberg, Germany}
\email{maltmann@ari.uni-heidelberg.de}

\begin{abstract}

We present the first entirely ground-based astrometric determination
of the proper motion for the Fornax Local Group Dwarf Spheroidal
satellite galaxy of the Milky Way, using CCD data acquired with the
ESO 3.5~m NTT telescope at La Silla Observatory in Chile. Our
unweighted mean from five Quasar fields in the background of Fornax,
used as fiducial reference points, leads to $\mu_\alpha \cos
\delta=0.62 \pm 0.16$~$\masy$, and $\mu_\delta=-0.53 \pm
0.15$~$\masy$. A detailed comparison with all previous measurements of
this quantity seems to imply that there is still no convincing
convergence to a single value, perhaps indicating the existence of
unnacounted systematic effects in (some of) these measurements. From
all available proper motion and radial velocity measurements for
Fornax, we compute Fornax's orbital parameters and their uncertainty
using a realistic Galactic potential and a Monte Carlo
simulation. Properties of the derived orbits are then compared to main
star formation episodes in the history of Fornax.

All published proper motion values imply that Fornax has recently
(200-300~Myr ago) approached perigalacticon at a distance of
$\sim$150~kpc. However, the derived period exhibits a large scatter,
as does the apogalacticon. Our orbit, being the most energetic,
implies a very large apogalactic distance of $\sim 950$~kpc. If this
were the case, then Fornax would be a representative of an
hypervelocity MW satellite in late infall.

\end{abstract}


\section{Introduction}

Studying the kinematics of the satellites of the Milky Way (MW) allows
us to adddress various fundamental issues such as the origin and
evolution of this satellite system (and the MW itself), the role of
tidal interactions in the evolution of the Local Group (LG) and the
matter distribution of the latter (including that of the dark matter,
thus allowing tests of some cosmological predictions
\citep{sha09}). This requires tracing their positions back in time, by
integrating their orbits, which, in turn, requires knowing their
current positions and their full 3-D space velocities. While radial
velocities for LG galaxies are known to better than
$\sim$5~km~s$^{-1}$, and their distances to $\sim$10\%, the biggest
source of uncertainty rests on their proper motions (PMs).

In the year 2000 we started a ground-based program aimed at
determining the absolute PM of three southern dwarf Spheroidal (dSph)
galaxies, Carina, Fornax and Sculptor, with respect to background
Quasars (QSOs), used as inertial reference points. Three to four
epochs of homogeneous CCD data were obtained using a single
telescope+detector setup over a period of eight years: Ours is the
first entirely optical CCD/ground-based PM study of an external galaxy
other than the Magellanic Clouds.

In M\'endez et al. (2010, hereafter Paper~I), we presented a detailed
description of our methods, as well as our first results for the PM of
Fornax, based on one QSO field (QJ0240-3434B). In this paper we
present our final PM for Fornax based on measurements from five QSO
fields in the background of Fornax. In Section~\ref{obs} we present a
summary description of our observational material, in
Section~\ref{pms} we explain how we obtained our PMs, and, finally, in
Section~\ref{comp} we compare our results to previous studies and
present our main conclusions.

The Fornax dSph galaxy, at a distance of 147~kpc (e.g.,
Pietrzy{\'n}ski et al. 2009), is relatively isolated, luminous and
well-resolved into individual stars. It seems to be dark matter
dominated, and has an estimated total mass of $\sim10^9$~M$_\odot$
(e.g., Walker et al. 2007). Unlike other dwarf galaxies, Fornax
harbours five globular clusters \citep{ho61}\footnote{Another notable
  exception is the tidally disrupted Sagittarius dwarf which, along
  with Fornax, are the most massive of the MW satellite galaxies after
  the LMC and SMC. Sagittarius appears to have globular clusters that
  are, or were, associated with it (e.g., Law \& Majewski 2010).}, and
it appears to have a complex stellar substructure in the form of
shell-like features indicative of recent merger activity (Coleman et
al. 2004, Coleman \& Da Costa 2005, Olszewski et
al. 2006). Furthermore, results by \citet{bat06} suggest that the
ancient stellar population in the centre of Fornax is not in
equilibrium (apparent as a non-Gaussian, double peaked velocity
distribution), which also points to a relatively recent accretion of
external material, such as gas accretion due to the merger with
another smaller stellar system.

\section{Observations}\label{obs}

All observations were carried out with the ``Superb Seeing Imager'',
SuSI2, attached to one of the Nasmyth focii of the ESO 3.5~m NTT
telescope at La Silla Observatory in Chile. The overall
characteristics of the detector are fully described in \citet{dod98a}
and \citet{dod98b}. We followed exactly the same observational,
reduction, and calibration procedures for all of our QSO fields as
described in Paper~I. Therefore, in as much as possible, our dataset
is homogeneous from this point of view (for certain limitations to
this statement, see Section~\ref{pms}). SuSI2 is a mosaic of two
2k$\times$4k EEV 44-82 chips (called \#45, on the West side, and \#
46, on the East side, for the adopted rotator angle of 0\degr). As
explained in Paper~I, we always placed our QSOs at a nominal pixel
position near $(x,y)=(3\,000,\,2\,100)$, close to the middle of chip
\#46, and only data from this chip was subsequently used for our
astrometry.

All of the astrometric observations, including those required to
compute the differential chromatic constants (DCR - see Paper~I,
Section~3.3.), were acquired through a Bessel R filter, whereas for
the blue frames needed to construct color-magnitude diagrams (CMDs) we
used the Bessel B filter, both of which are part of the standard set
of filters available for SuSI2.

Our initial list of QSOs in the background of Fornax comprised all
eleven distinct QSOs (one, reported in Paper~I, forms a gravitational
lens pair) reported by \citet{tin97} (their Table~5) and \citet{tin99}
(his Table~1). Unfortunately, several of them proved to be useless for
astrometry: deep, good-seeing (FWHM$\sim$0.5~arcsec) images taken
during the first epoch revealed that they were either too faint for
our required astrometric $S/N$ ($>200$ integrated over the PSF fitting
radius; see Paper~I, Section 3.2 and Figure~4), had a noticeably
elongated structure (and therefore had a different, usually more
extended, PSF than that of the stars), had a very nearby (usually
stellar-like) bright companion, or were definitely blended with field
stars. As explained at length in Paper~I, any of these features render
these targets unsuitable for high-accuracy relative astrometry. These
problematic QSOs were dropped from the observing list in subsequent
epochs, and we concentrated on ``clean'' (as far as we could determine
with a seeing of $\sim$0.5~arcsec) QSOs. Our final list of five QSOs,
from which we were able to determine the PM of Fornax, is given in
Table~\ref{tbl1}, where we also show the full list of known QSOs
behind Fornax from the above-cited references.

\begin{figure}
\epsscale{.80}
\plotone{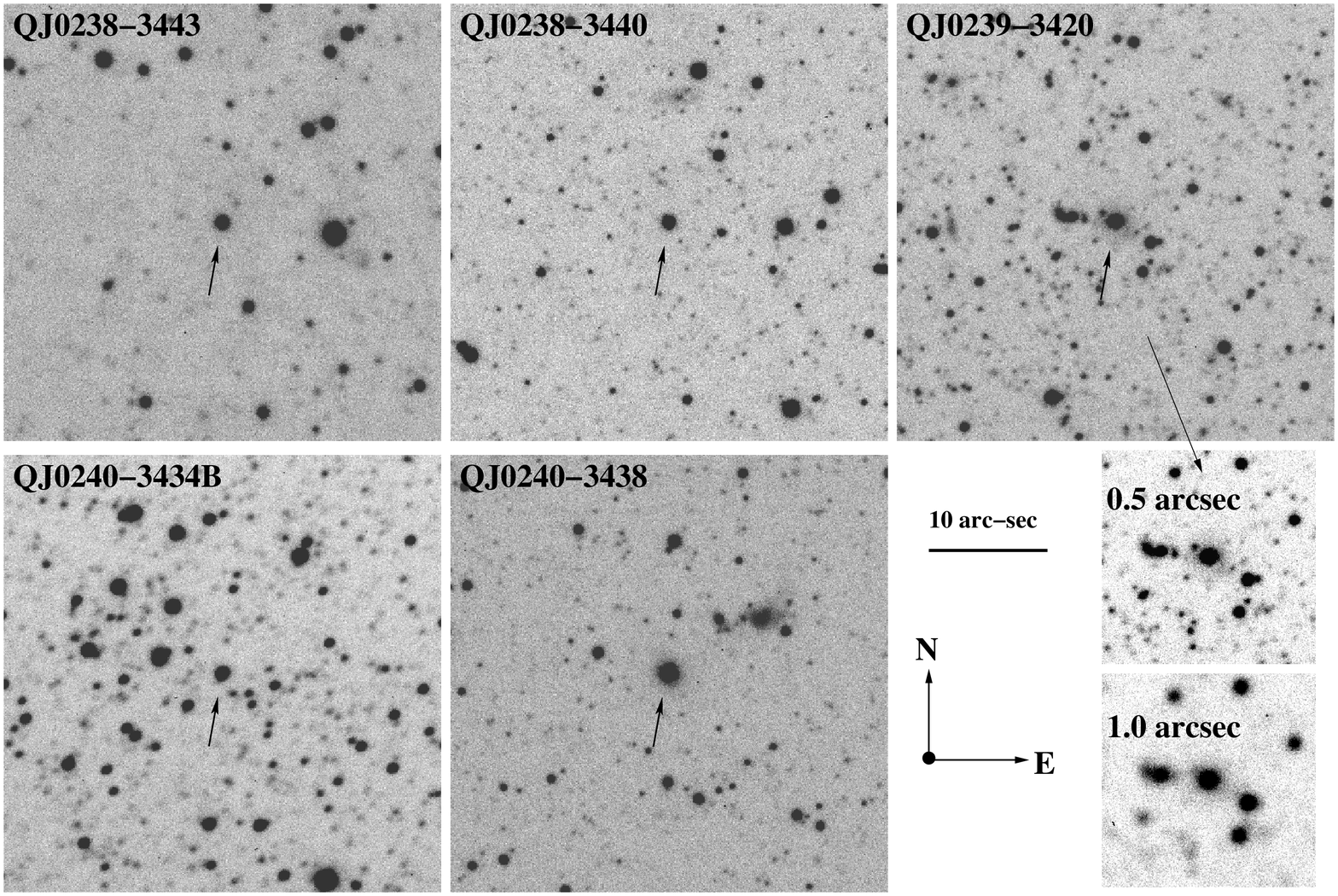}

\caption{Finding charts indicating the configuration and surroundings
  of our five QSO fields. The QSOs are indicated by an arrow. The
  field-of-view is $\pm$250~pix around the QSO (approximately
  40~arcsec on a side), and the FWHM of all images is
  $\sim$0.5~arcsec. With the exception of QJ0239-3420, all QSOs are
  well isolated from surrounding sources (see text). Our tests
  indicate that the small source to the NW of QJ0239-3420 does not
  affect the astrometry.  To illustrate the effect of seeing, in the
  bottom right inset, we show a zoom of the area around QJ0239-3420
  for best (top) and worse (bottom) FWHM frames for this
  field. \label{fields}}

\end{figure}

In Figure~\ref{fields} we show the stellar configuration in the
immediate surroundings of each of our five selected QSOs. All of these
images were acquired on the first epoch, when we consistently had very
good seeing (FWHM$\sim$0.5~arcsec). The only potentially problematic
case was that of QJ0239-3420 which has a faint companion to the NW, at
a distance of $\sim1.4$~arcsec. Fortunately, this nearby source is
very dim (with a Star/QSO peak brightness ratio smaller than 0.1), and
far enough (farther than $1.2\times FWHM$, adopted as the PSF fitting
radius, see Paper~I, Section~3.2 and Figure~4) that, even on our worse
frames (with a FWHM of $\sim1.0$~arcsec), it did not pose a problem
for the astrometric solution, and so it was fully included in our
analysis below.

Table~\ref{tbl2} contains a summary of the observational material that
was, in the end, used to compute our PMs. We note that more data was
acquired for these QSO fields, but proved to be useless on account of
bad seeing, poor image quality (deteriorating the astrometric
solution), or bad sky transparency. As explained in Paper~I, at the
start of our program we adjusted the exposure time on the basis of
seeing (typical values were between 300~s and 900~s), but later, it
was decided to use a fixed integration time of 900~s in all cases for
simplicity.

\section{Proper motions} \label{pms}

\begin{figure}
\epsscale{.52}
\plotone{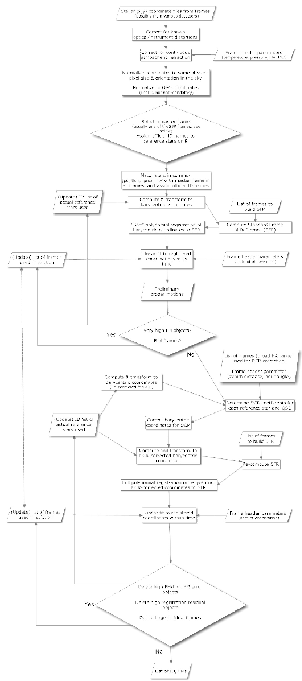}

\caption{Flow diagram from raw stellar pixel coordinates to final PMs
  for the QSO and bona-fide Fornax reference stars, based on
  procedures detailed in Paper~I. We note that there is an elaborate
  previous selection of ``suitable'' (good $S/N$, isolated) reference
  stars (potential members of Fornax) from which precise pixel
  coordinates are derived through PSF fitting. All these steps ocurr
  prior to the entry point in this flow diagram, and are fully
  explained in \citet{cos09} and Paper~I.\label{flow}}

\end{figure}

To determine the PMs we used the same procedure described at length in
Paper~I. A flow-chart summary of the full process is presented in
Figure~\ref{flow}. We refer the reader to Paper~I for further details
on the methodology and precise meaning of all steps. Even though in
our PM solutions we only included frames acquired within
$\pm1.5$~hr. from the meridian (see Table~\ref{tbl2}), all of our
coordinates were (pre)-corrected for (continuous) atmospheric
refraction and DCR as described in Paper~I. We also excluded from the
PM solution all frames with a $FWHM > 1.0$~arcsec, as they clearly
deteriorated the linear fit of Baricentric position {\it vs.} epoch
diagram (see, e.g., Fig.~12 in Paper~I), where these frames stand-out
due to their large scatter.

All PM data were treated as homogenously as possible, including the
following constraints:

\begin{itemize}

\item All stars with $\mu>2.0$~$\masy$ in our initial local reference
  system (LRS, basically a set of ``high-quality'' reference stellar
  images, eventually bona-fide Fornax stars - see Paper~I), indicating
  that they are either foreground Galactic stars, or that they have a
  high (pseudo)-PM, possibly due, e.g., to a faint unresolved
  companion or another problem in the image), were purged from the
  LRS. The full geometric registration (and PMs) for the remaining LRS
  stars and the QSO were re-computed in an iterative process,

\item All LRS stars that exhibited a registration residual $\ge
  3\sigma$ of the formal rms of the two-dimensional geometric
  registration in the $X$ or $Y$ coordinates in at least four (not
  necessarily consecutive) frames, were eliminated from the LRS. For
  registration we used a full 3$^{rd}$ order polynomial fit, which has
  been justified in Paper~I,

\item All LRS stars with $\sigma_{\mu} > \sigma_{\mu_c}$~$\masy$ were
  excluded from the LRS. The cut value, $\sigma_{\mu_c}$, was
  estimated for each field based on the distribution of PM errors {\it
    vs.} magnitude (see Table~\ref{tbl3}),

\item LRS stars exhibiting a PM error larger than that of the bulk of
  the LRS stars at a given magnitude (even if they, individually, had
  $\sigma_{\mu} < \sigma_{\mu_c}$) were eliminated. This cut was done
  visually, on plots of PM error vs. magnitude for each QSO field,

\item Finally, we expunged objects that in the CMD appeared not to
  belong to Fornax. This photometric cleansing was done by first
  calibrating our instrumental photometry following the procedure
  described in Paper~I (Section~4), and then by comparing our
  resulting CMD with that of \citet{ste97}, which defines the main
  features of the Fornax CMD.

\end{itemize}

\begin{figure}
\epsscale{1.1}
\plotone{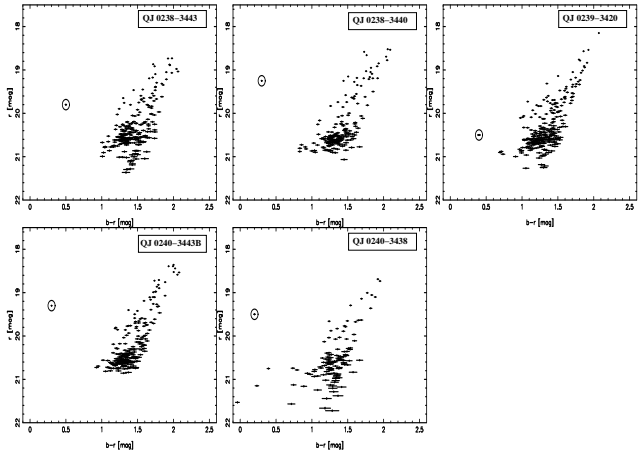}

\caption{{\it Instrumental} CMDs for our five QSO fields. The QSO is
  idicated by an ellipse. For comparison purposes among the different
  fields, our PSF photometry has been {\it approximately} calibrated
  by adopting a red-clump magnitude and color of $R\sim20.6$ and
  $(B-R) \sim 1.3$ respectively (from Stetson~1997).\label{cmd}}

\end{figure}

In Table~\ref{tbl3} we show a summary of what results from applying
the above criteria to the initial set of LRS stars for each of our
five QSO fields. After applying all previously described cuts, we
verified that we ended with a uniform $(X,Y)$ distribution of the LRS
stars (see, e.g., Figure~15 in Paper~I), and with a reasonable
distribution in magnitude and color. The resulting CMDs for the LRS
stars and the respective QSO are shown in Figure~\ref{cmd}. We note
that in Paper~I we calibrated our photometry approximately by adopting
a color of $(B-R) \sim 1.3$ for the red-clump from \citet{ste97} and
the QSO blue magnitude from the works by \citet{tin97} and
\citet{tin99}. However, we noticed that, probably as result of
uncertainties in the blue photographic magnitudes for the QSOs (and/or
possible QSO variability), the ordinate in these figures had a large
zero-point variation from field to field, and hence the red-clump did
not fall at the same apparent $R$ magnitude for the five fields, as is
expected. We therefore decided to adopt a calibration based on the
color of the red-clump (as before), but fixing, instead, the magnitude
of the red-clump at $R\sim20.6$, also from the photometry by Stetson
(1997, his Figure~7).

\begin{figure}
\epsscale{1.1}
\plotone{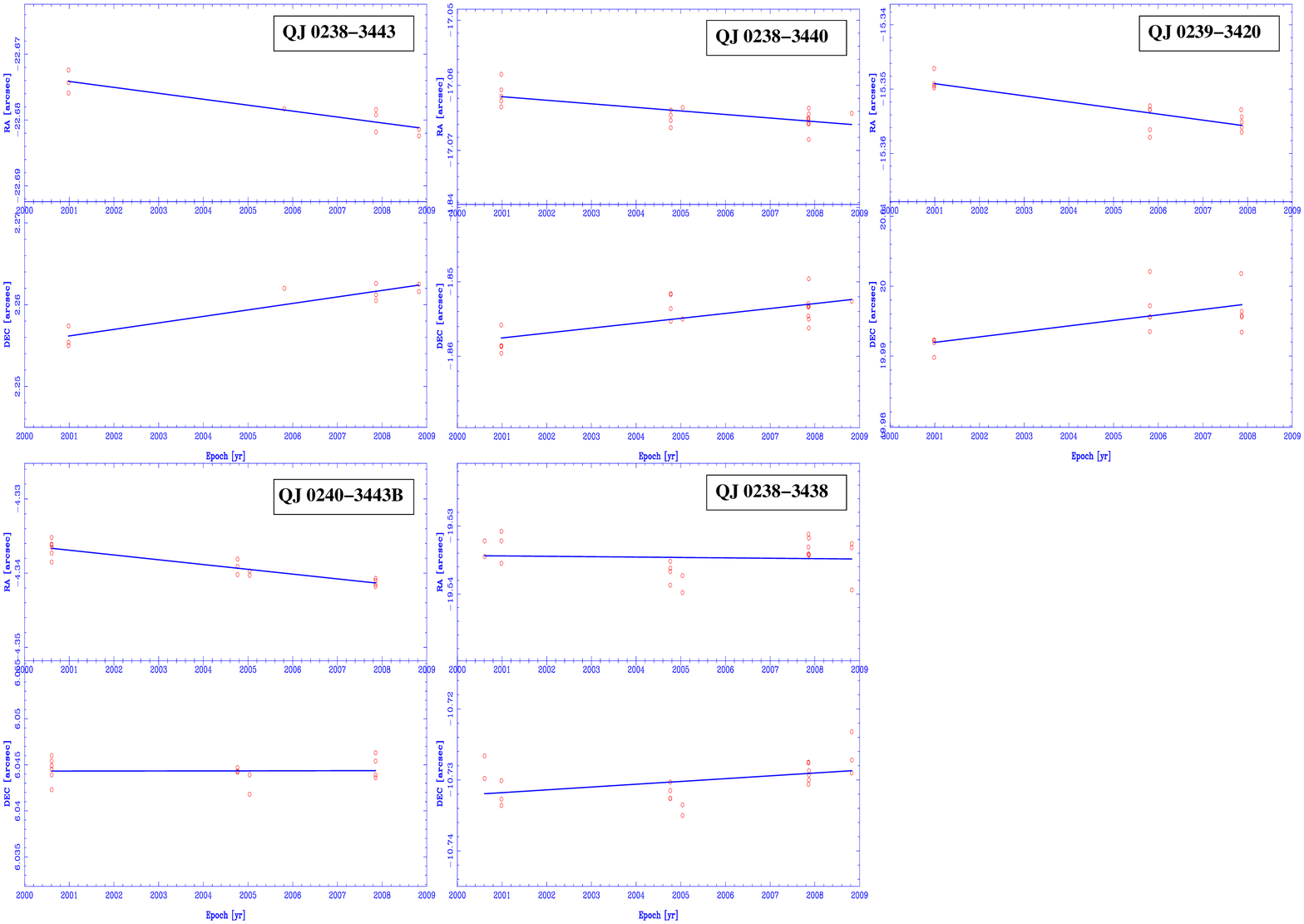}

\caption{QSO barycentric position with respect to the (final) LRS
  stars {\it vs.}  epoch for our five QSO fields. The lines are
  (unweighted) fits to the data points, and the negative of the slope
  of these fits corresponds to the PM of Fornax, whose values are
  tabulated in Table~\ref{tbl4}.\label{posepoch}}

\end{figure}

\begin{figure}
\epsscale{1.1}
\plotone{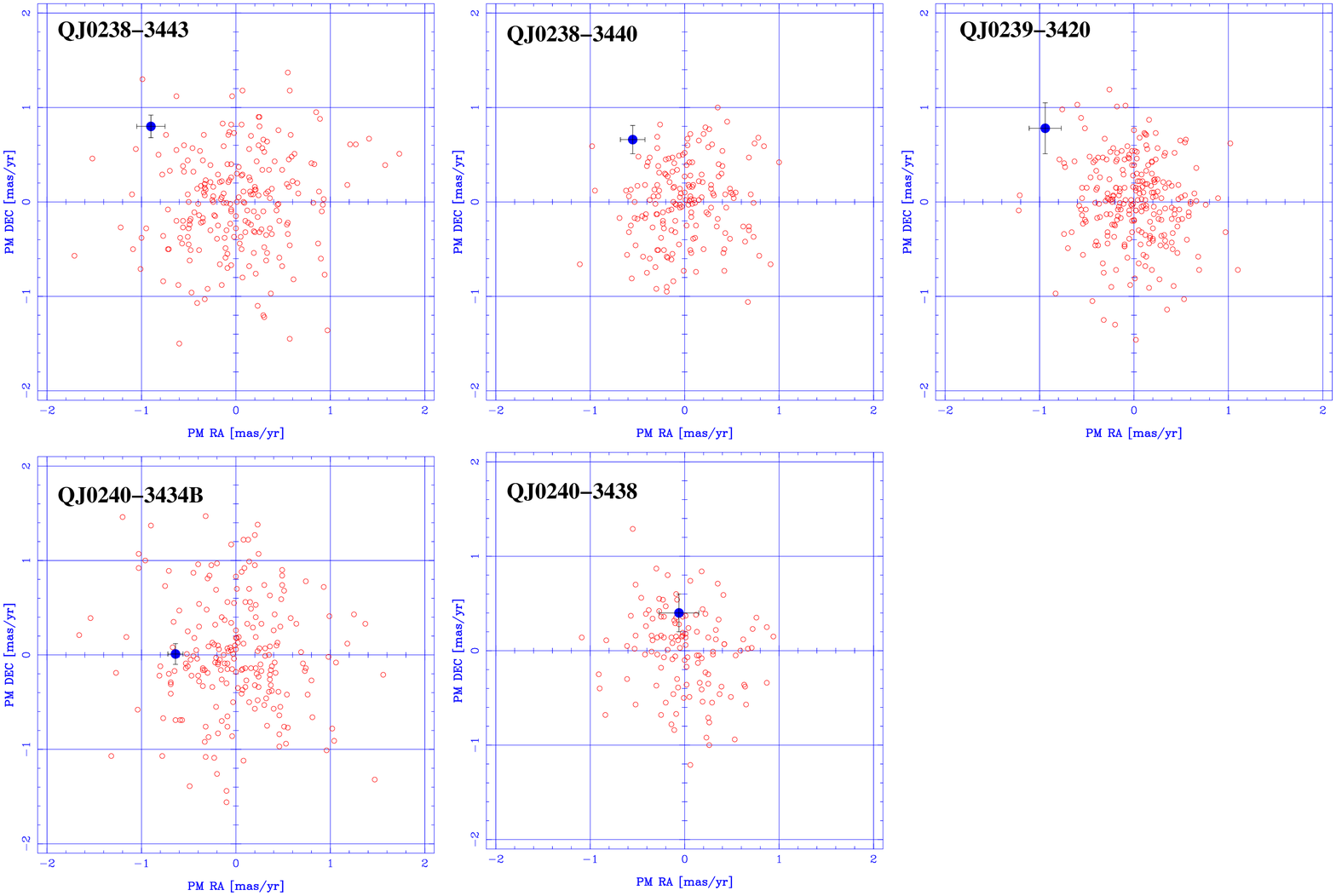}

\caption{Vector-point diagrams for our five QSO fields, after having
  applied the selection criteria described in the text. The QSO is
  indicated by a solid dot with error bars ($1\sigma$ uncertainty on
  the slope of the barycentric position {\it vs.} epoch fit). The
  (pseudo-) PM error of individual LRS stars is similar to that of the
  QSO, so the mean motion of the bulk of the LRS stars has a very
  small uncertainty, and our final error is dominated by the PM
  uncertainty of the QSO itself (see Costa et al. 2011).\label{vpd}}

\end{figure}

In Figures~\ref{posepoch} and~\ref{vpd} we show, respectively, the
barycentric position {\it vs.} epoch and the vector-point diagrams for
our five fields, for the final LRS stars and the corresponding
background QSO.

After processing all of the QSO fields following the protocols
described in the previous paragraphs, we computed PMs for the five QSO
fields presented in the upper part of Table~\ref{tbl1}. The results
are summarized in Table~\ref{tbl4}, and are plotted in
Figure~\ref{pmind}. Table~\ref{tbl4} gives the PM in RA and DEC as
well as the overall rms of the linear fit of the barycentric
coordinates {\it vs.}  epoch. We recall that the slope of this linear
fit gives us the PM of the Fornax stars with respect to the QSO, as
explained in Paper~I.  Figure~\ref{pmind} clearly shows that our five
measurements scatter more widely than would be expected from their
error bars, and therefore computing a weighted mean does not seem
appropriate for these data. Instead, and not knowing the source of the
additional uncertainty affecting our proper motion measurements, we
opted for computing an unweighted mean over the resulting values for
all five QSO fields. From Table~\ref{tbl4} we thus arrive at the
following values for the PM of the Fornax dSph galaxy (and the
standard deviation of the mean): $\mu_\alpha \cos \delta=0.62 \pm
0.16$~$\masy$, and $\mu_\delta=-0.53 \pm 0.15$~$\masy$.

\subsection{Comparison to other studies} \label{comp}

Only two astrometric determinations of the PM for the Fornax dSph
galaxy are available, namely that by \citet{din04}, based on a
combination of ground-based plates and Hubble-WFPC data, and that
based exclusively on HST data by Piatek et al. (2007, hereafter PI07),
who present revised values to those reported earlier (pre-CCD Charge
Transfer Inefficieny corrections) in \citet{pia02}.

\begin{figure}
\epsscale{.90}
\includegraphics[angle=-90,scale=.90]{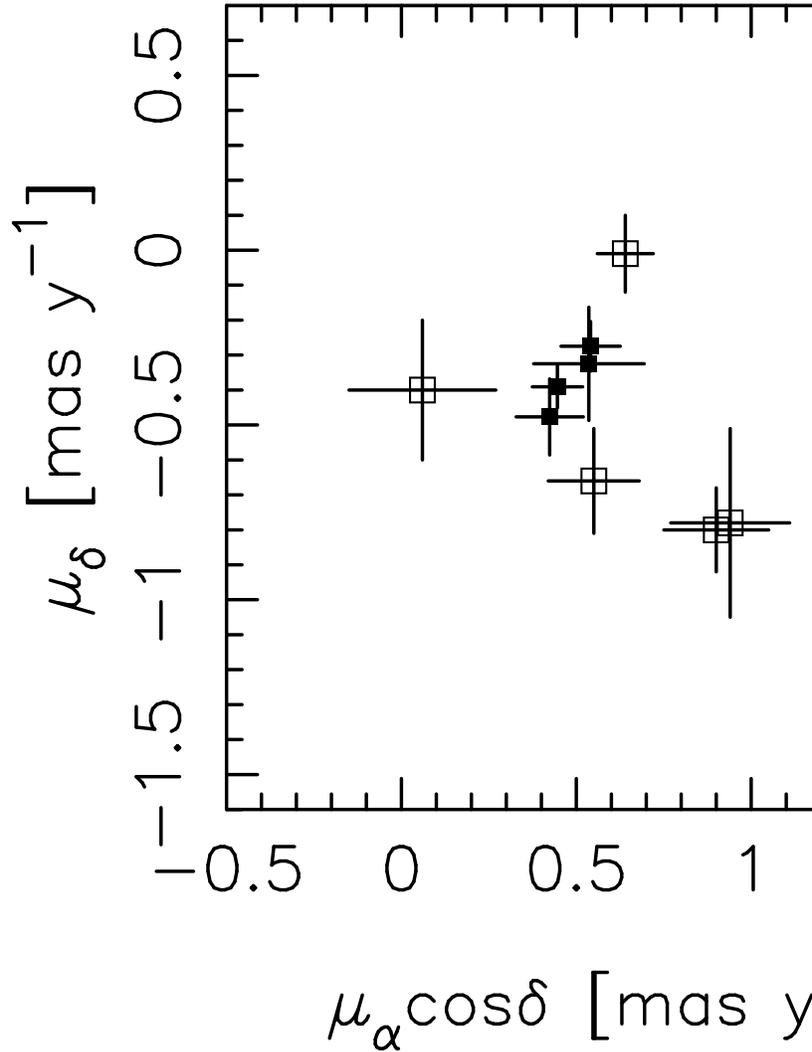}

\caption{Individual Fornax PMs determined from our five QSO fields
  (from Table~\ref{tbl4}, open squares). Filled squares are the HST
  PC2+STIS results from four QSOs by Piatek et al. 2007, their
  Table~3. Three of these QSOs are common to ours (see our
  Table~\ref{tbl5}). \label{pmind}}

\end{figure}

Figure~\ref{pmind} compares our results from individual QSOs, to those
obtained by PI07. We have three QSO fields in common with PI07; their
values, along with our measurements, are given in
Table~\ref{tbl5}. From this Table we see that the difference of the
mean PMs derived from the QSOs in common between these two studies is
less than $1 \sigma$ of our error in the mean in both, RA and DEC. On
the other hand, the overall weighted mean value reported by PI07,
based on 4 QSOs (their Table~3) of $\mu_\alpha \cos \delta=0.476 \pm
0.046$~$\masy$, and $\mu_\delta=-0.360 \pm 0.041$~$\masy$ differs by
about $1 \sigma$ from our overall unweighted-mean PM values. Even
though our results are not statistically inconsistent with those of
PI07, it is apparent from Figure~\ref{pmind} that our individual
results do exhibit a much larger scatter than their results
($\sigma_{\mu_\alpha \cos \delta} = 0.36$~$\masy$ and
$\sigma_{\mu_\delta} = 0.34$~$\masy$ vs. $\sigma_{\mu_\alpha \cos
  \delta} = 0.06$~$\masy$ and $\sigma_{\mu_\delta} = 0.09$~$\masy$ for
PI07). We have no explanation for these differences.

\begin{figure}
\epsscale{.90}
\includegraphics[angle=-90,scale=.90]{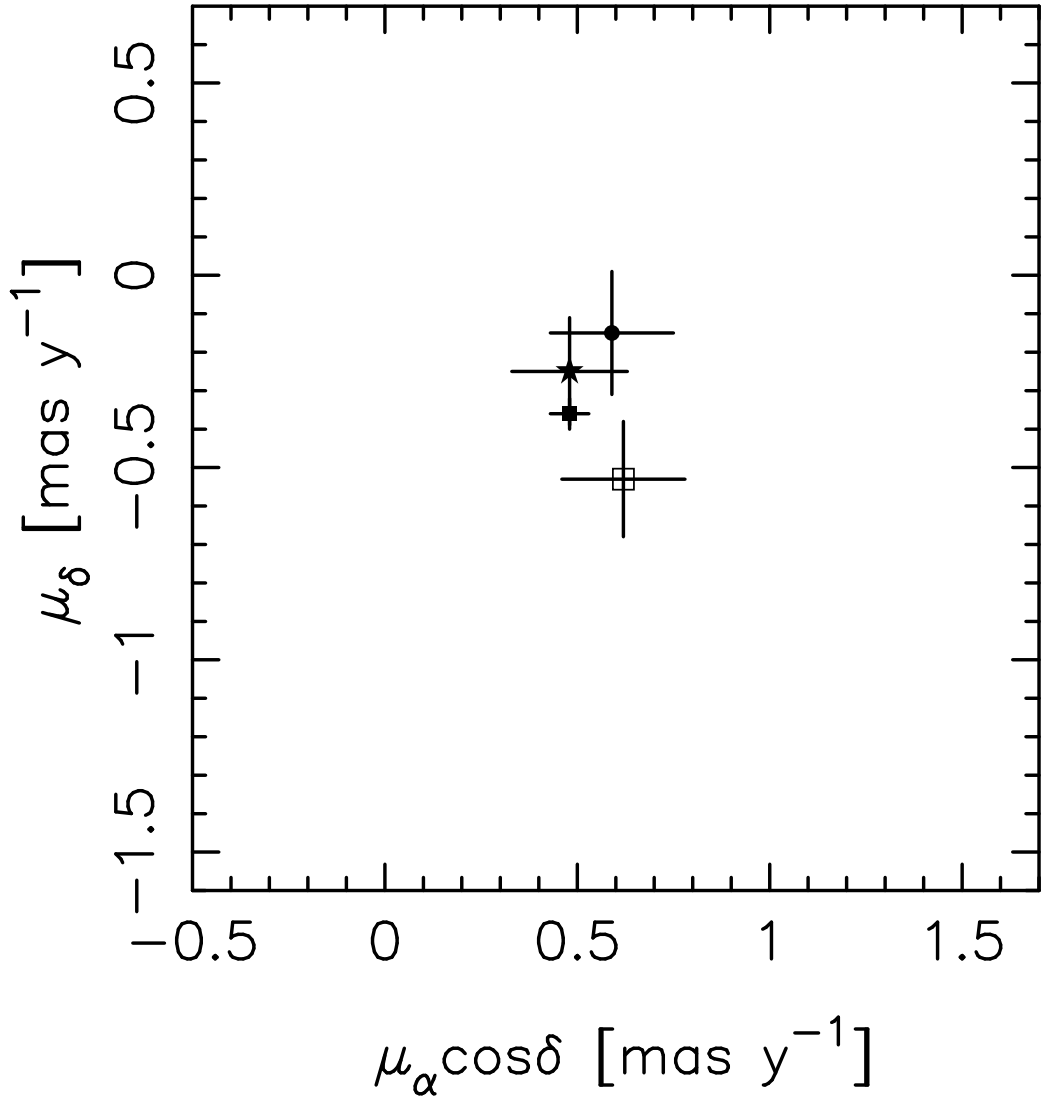}

\caption{Mean Fornax PM and their $1\sigma$ errors as determined by
  various authors: Filled dot from \citet{din04}, filled square from
  \citet{pia07}, filled star from \citet{wal08}, and open square this
  work (see Table~\ref{tbl6}).\label{pmwei}}

\end{figure}

The PM values published so far, along with our own value are given in
Table~\ref{tbl6}. More recently, \citet{wal08} have used a
non-astrometric method called ``perspective rotation'' which is also
included in the table (for details about this method see Paper~I),
because it provides a completely independent measurement of the
tranverse motion of Fornax (albeit with a larger error than the more
recent astrometric determinations). All of these values are plotted,
for comparison, in Figure~\ref{pmwei}. From the plot and figure we can
see that, in general, there is a good agreement between all
measurements; indeed, none of them depart by more than $2 \sigma$ from
the straight average of the values in Table~\ref{tbl6}, with our
result being however the most extreme in this sense. Also, averaging
the results from all of the authors, we find that the PM in DEC shows
a larger scatter ($\sigma_{\mu_\delta} = 0.16$~$\masy$) than that in
RA ($\sigma_{\mu_\alpha \cos \delta}=0.064$~$\masy$). This would
suggest that (some of) these measurements are affected by unaccounted
systematic effects. Of the obvious culprits, we can mention that
ground-based astrometric data are affected by DCR. In the case of
Fornax data (DEC$\sim -34\degr$) acquired from the Southern
Hemisphere, DCR should however mostly affect RA PMs (we stress that
our data have been corrected for this effect as far as possible, see
Paper~I).

On the other hand, HST data, while not affected by DCR, can be
affected by ``Charge Transfer Inefficiency'' (CTI) in the CCD
detectors due to the very low sky background which is insufficient to
fill-in the empty charge traps on the detector. These traps evolve in
time because they are produced by in-flight radiation damage, and
induce systematic (time dependent) position shifts in the detected
sources (for details see, e.g., \citet{bri06}, especially his
Figure~10). HST data has also been corrected for this effect, again,
as best as they could (compare, e.g., Table~3 from \citet{pia02}
(CTI-uncorrected) and PI07 (CTI-corrected)). We note however that,
since the HST cameras were not oriented exactly along DEC in the
parallel readout direction, which is the direction mostly affected by
CTI (see, e.g., Figures~2, 3 and~4 in PI07), it is difficult to
ascribe the DEC scatter to this problem. Also, as mentioned
previously, HST data shows a much smaller intrinsic scatter than our
measurements (Figure~\ref{pmind}), thus suggesting small remaining
systematics due to CTI.

\section{Discussion and outlook}

One of the ultimate goals of these studies is a determination of a
reliable orbit for Fornax. It is interesting to compare the impact of
the different values given in Table~\ref{tbl6} on the kinematical and
orbital parameters that can be derived from these
measurements. Table~\ref{tbl7} gives the Heliocentric PMs and the
corresponding tangential velocities in Galactic coordinates for the
different values reported in Table~\ref{tbl6}. From this table, it is
clear that our PM value will render one of the most energetic orbits
yet derived for Fornax.

For illustration purposes, and also as a key ingredient to compute and
interpret the orbit of Fornax from the above motions in Galactic
coordinates, we have computed velocities in the Heliocentric (HC) and
Galactocentric (GC) reference systems. For a detailed description of
these various reference systems, and the equations relating them, the
reader is referred to \citet{cos09}. They are shown in
Table~\ref{tbl8}, along with model-independent kinematical and orbital
parameters derived from the PM values taken from different
studies. Given the large values of $V_{t_{\mbox{GC}}}$ in comparison
with the $V_{r_{\mbox{GC}}}$ derived from all PM measurements, it is
clear that Fornax must be close to perigalaticon at this time. Also,
given the rather small ratio of $L_z/L$ it is clear that there must be
significant excursions of Fornax away from the Galactic plane,
regardless of the adopted PM values. We note again that, of all
measurements available, our PM measurement yields the most extreme
orbit.

\begin{figure}
\epsscale{.90}
\includegraphics[angle=-90,scale=.70]{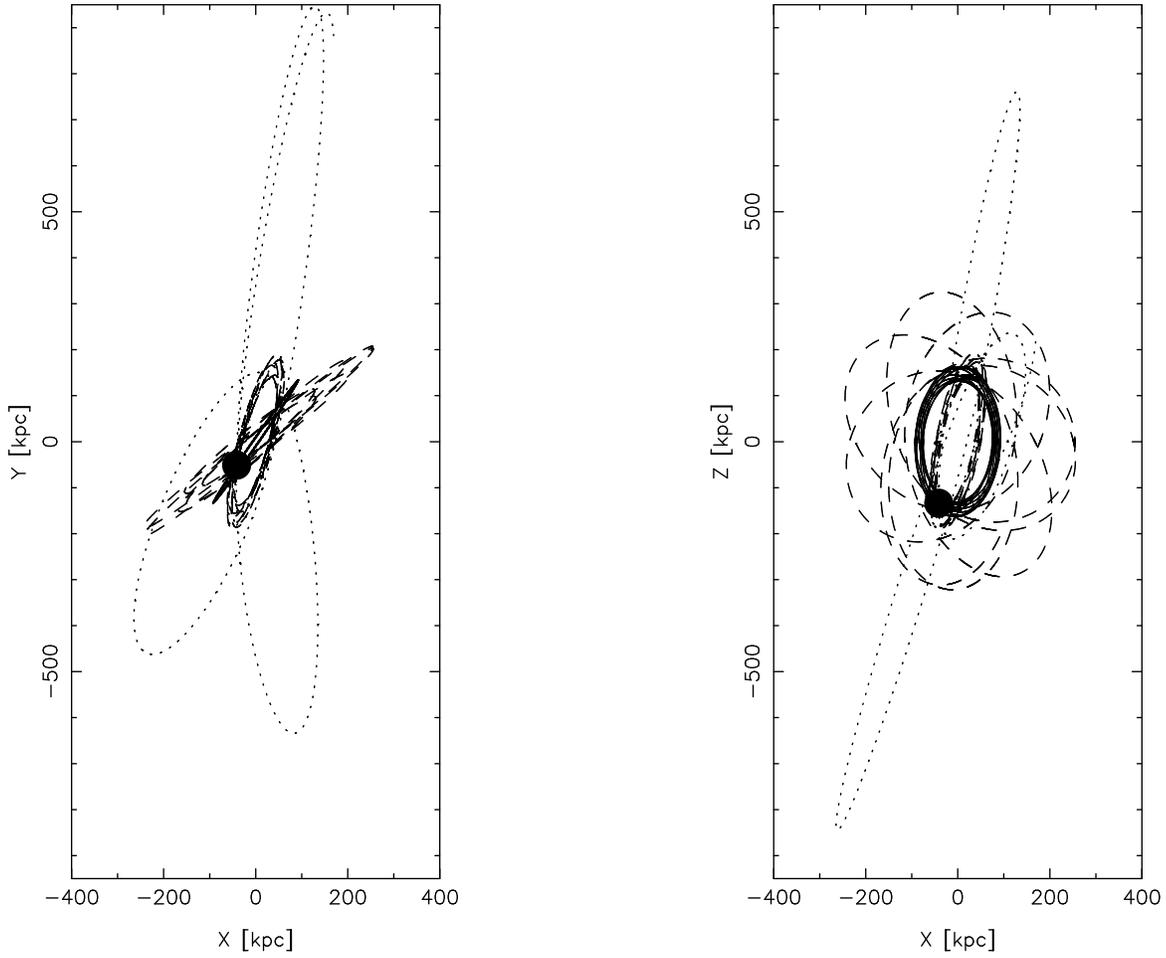}

\caption{X-Y and X-Z projections of the orbit for Fornax from PM
  values by different authors: Dashed line is \citet{din04},
  dot-dashed \citet{pia07}, full line \citet{wal08}, and dotted line
  this work. The filled dot at $(X,Y,Z)=(-41.4,-50.9,-133.9)$~kpc
  indicates the current position of Fornax.\label{xyz}}

\end{figure}

\begin{figure}
\epsscale{.90}
\includegraphics[angle=-90,scale=.70]{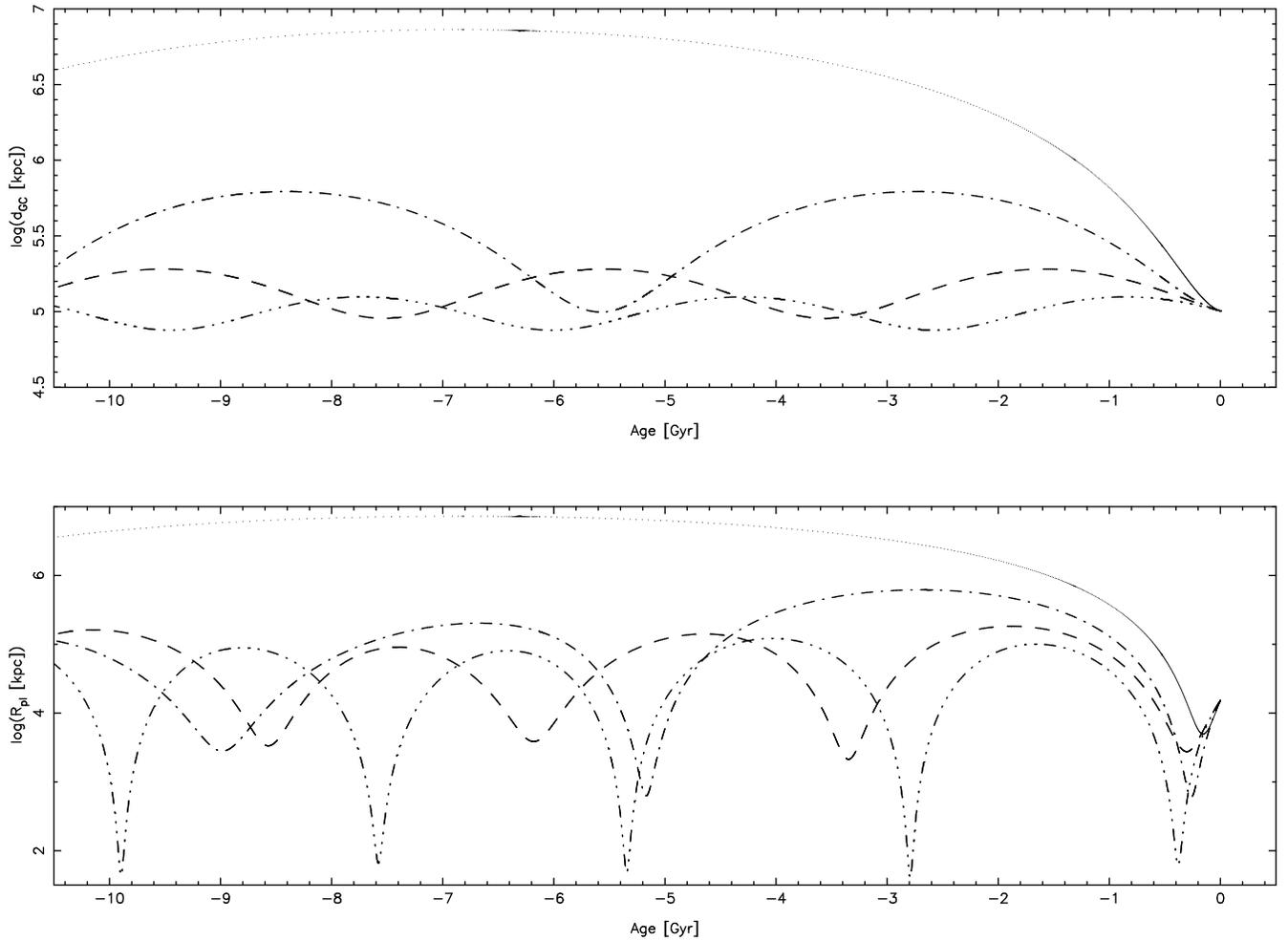}

\caption{Fornax distance (log scale) as a function of age from our
  orbit integrations, for PM values from different authors. Top panel
  is for the Galactocentric distance, bottom panel is for the distance
  projected on the plane of the MW. Both panels: Dot-dashed line is
  for \citet{din04}, dashed line for \citet{pia07}, double dot-dashed
  line for \citet{wal08}, and dotted line for this work.\label{dgct}}

\end{figure}

We have computed Galactic orbits for Fornax by integrating back in
time the equation of motion under a realistic three-component (disk,
halo and spheroid) model Galactic potential \citep{jo95}, and using
the current position and velocity (and their uncertainties, in a Monte
Carlo scheme) as initial conditions. Details of the integrator, and
the Monte Carlo simulations, are discussed in \citet{din04}. A
particularly important feature of the integrator used is its care to
conserve energy and total angular momentum, obvious features that are
howevever sometimes tricky to achieve over the full orbit,
particularly when the number of integration steps ($\sim$several
thousands in our case) is large. The adopted gravitational potential
is strictly axi-symmetric, and it does not include the Galactic bar,
nor the MW spiral pattern. Recent numerical simulations by
\citet{all08}, that include these non axi-symmetric components,
indicate however that their effect, in particular in the orbits of
Galactic globular clusters (and therefore also in the case of the more
external satellite galaxies), is minimal, and it should not alter our
conclusions importantly. The results of these integrations are shown
in Table~\ref{tbl9} and in Figure~\ref{xyz}. In Table~\ref{tbl9} the
meaning of the different columns is rather obvious, except perhaps for
the last two columns that correspond to the crossing (vertical)
velocity of Fornax through the Galactic plane ($V_z$) and the total
speed at perigalacticon ($V_p$). As it can be seen, regardless of what
PM values one adopts, comparing the $V_p$ values to the
$V_{t_{\mbox{GC}}}$ in Table~\ref{tbl8} confirms that the current
Fornax position lies indeed near perigalacticon (see also
Figure~\ref{dgct}, top panel). Actually, all PMs indicate that the
minimum distance, projected into the Galactic plane, happened 200 to
300~Myr ago (Figure~\ref{dgct}, bottom panel). This is a very
interesting quantity, because \citet{bat06} have found evidence of at
least three distinct stellar components in Fornax: a young population
(few 100~Myr old) concentrated in the centre of the galaxy, visible as
a Main Sequence in the CMD; an intermediate age population (2-8~Gyr
old, possibly related to a shell structure in Fornax, described in the
next paragraph); and an ancient population ($>10$~Gyr). More recently,
\citet{kir11} have also found (see, e.g., their Figure~2) evidence of
enhanced star formation in the range 200-300~Myr. One could therefore
conclude that the latest episode of star formation on Fornax may have
been indeed triggered by its perigalacticon passage.

When tracing the galaxy back in time, the uncertainty in the computed
position increases as time goes on, for a given uncertainty in the
initial conditions, and, as we extrapolate further back in time,
different PM values lead to quite different orbits. This is clearly
shown in Figure~\ref{dgct}, which shows the distance of Fornax from
the Galactic center ($d_{\mbox{GC}}$, top panel) and the same distance
projected onto the Galactic plane ($R_{\mbox{pl}}$, bottom panel) as a
function of time from now. Initially all PM values produce a similar
$d_{\mbox{GC}}$, $R_{\mbox{pl}}$ {\it vs.} time, but the solutions
diverge afterwards. As mentioned before, our value renders the most
extreme solution, basically indicating that in a Hubble time Fornax
has not completed an orbit yet. This is at odds with all of the other
solutions which, while being different among themselves, do indicate
nevertheless several perigalacticon passages in the last 10~Gyr. If,
as argued before, perigalacticon has had an influence on enhancing
star formation, the stellar population results by \citet{bat06} favour
a rather long orbital period, thus supporting our solution. We note
here that the 2~Gyr population belonging to the shell structure found
by \citet{col04}, \citet{col05}, and \citet{ols06} has been
interpreted by these authors as the product of a merger with a
smaller, gas-rich system, and may not bear any relation to the
interaction between Fornax and the MW. We are thus seemingly left with
two significant star formation episodes, one at (200-300~Myr), close
to perigalacticon, and another one at age~$>10$~Gyr, which again
argues for a longer orbital period (interestingly, from
Figure~\ref{dgct} we see that $\sim2$~Gyr mark the most recent
apogalactic position for Fornax for the PI07 and \citet{wal08}
orbits). Our extended orbit implies that Fornax would belong to one of
the ``hypervelocity'' satellites of the MW, as argued by
\citet{kal06}, \citet{bes07}, \citet{pia08} (results all based on the
same HST data set).
One could perhaps wonder that the late infall nature of Fornax's
orbits (assuming our PMs) is possibly related to the fact that it is
the only satellite dwarf galaxy of the MW (Mateo 1998), along with the
spatially dissipated Sagittarius dwarf (Law \& Majewski 2010), that
harbour a globular cluster population.

A final word of caution regarding the above discussion is that, from
Table~\ref{tbl9}, we see that current PM measurements for Fornax
implies that the range of derived orbital parameters is quite broad,
and it seems adventurous to extract strong conclusions from them. As
already mentioned, of particular concern is the possible existence of
yet unaccounted systematic effects in the PM measurements. Only
high-accuracy future astrometric satellite measurements, like GAIA,
with expected uncertainties of a few $\mu$arc-sec (average over many
stars) could help resolve this issue.

%
%

\clearpage


\acknowledgments

RAM and EC acknowledge support by the Fondo Nacional de
Investigaci\'on Cient\'{\i}fica y Tecnol\'ogica (Fondecyt project
No. 1070312), the Chilean Centro de Astrof\'{\i}sica (FONDAP project
No. 15010003) and the Chilean Centro de Excelencia en Astrof\'{\i}sica
y Tecnolog\'{\i}as Afines (PFB 06). MHP acknowledges support by
Project \# 4721-09 from Universidad de Tarapac\'a. CG acknowledges
support by the Instituto de Astrof\'isica de Canarias (P3-94) and by
the Ministry of Education and Research of the Kingdom of Spain
(AYA2004-06343). RAM acknowledges support on various stages of
reduction process by Dr. Matias Radiszcz. We are greatful to the ESO
OPC for their continued support of this long-term program, as well as
to the La Silla Scientists, Engineers and Operations staff for their
continuous help in the course of the program, especially Dr. Michael
F. Sterzik and Mr. Federico Fox. We are also greatful to Dr. Dana
Casseti-Dinescu, who ran the orbit simulations for our data using her
code. We are also very gratefull to an anonymous referee for many
useful comments.



{\it Facilities:} \facility{ESO-NTT (SuSI2)}.


\clearpage


\clearpage


\clearpage


\clearpage


\clearpage


\clearpage


\clearpage


\clearpage


\begin{deluxetable}{ccccccc}
\tabletypesize{\scriptsize}
\rotate
\tablecaption{QSOs in the background of Fornax, from \citet{tin97} and \citet{tin99}, in
  order of increasing RA\tablenotemark{a}.\label{tbl1}}
\tablewidth{0pt}
\tablehead{
\colhead{QSO ID} & \multicolumn{2}{c}{RA    (J2000)    DEC} & \colhead{B} & \colhead{$\rho$\tablenotemark{b}} &
\colhead{$\theta$\tablenotemark{b}} & \colhead{Comments} \\
       & [hh:mm:ss.s] & [$^0$:$'$:$''$] & [mag] & [arcmin] & [deg]      &
}
\startdata
QJ0238-3443  & 02:38:43.8 & -34:43:53 & 20.2 & 22.9 & 222 \\
QJ0238-3440  & 02:38:55.6 & -34:40:45 & 20.1 & 19.0 & 223 \\
QJ0239-3420  & 02:39:49.0 & -34:20:00 & 20.6 & 07.3 & 344 \\
QJ0240-3434B & 02:40:08.2\tablenotemark{c} & -34:34:22\tablenotemark{c} & 19.9 & 07.6 & 166 & Reported in Paper~I\\
QJ0240-3438  & 02:40:38.7 & -34:38:58 & 20.2 & 14.5 & 146 \\
\hline
QJ0239-3425  & 02:39:32.9 & -34:25:25 & 20.4 & -- & -- & No finding chart available \\
QJ0239-3421  & 02:39:36.9 & -34:21:30 & 19.9 & -- & -- & Fuzzy, elongated QSO image, with a FWHM 48\% larger than that of stars in the FOV \\
QJ0240-3434A & 02:40:07.7\tablenotemark{d} & -34:34:20\tablenotemark{d} & 19.1 & -- & -- & Blended to foreground star \\
QJ0240-3437  & 02:40:19.0 & -34:37:20 & 17.9 & -- & -- & QSO exhibits complex structure \\
QJ0241-3420  & 02:41:57.9 & -34:20:49 & 20.4 & -- & -- & No finding chart available, far from Fornax's center \\
QJ0242-3426  & 02:42:06.5 & -34:26:12 & 21.8 & -- & -- & Faint and far from Fornax's center \\
QJ0242-3424  & 02:42:19.9 & -34:24:20 & 21.5 & -- & -- & Faint and far from Fornax's center \\
\enddata

\tablenotetext{a}{The upper part of the table shows the QSOs used in
  this work. The bottom part indicates those which could not be used
  for astrometric purposes or were otherwise discarded for various
  reasons, as indicated under column ``Comments''.}

\tablenotetext{b}{To compute the angular distance ($\rho$) and
  position angle ($\theta$) of the QSO with respect to Fornax, we have
  taken the J2000 coordinates for the center of Fornax from Table~1 of
  \citet{mat98}, namely RA=02:39:59, and DEC=-34:27.0. We note that
  these values differ by more than 1~arcmin in RA and 3~arcmin in DEC
  from the values presented in the review of LG galaxies by van den
  Bergh (1999, Table~1) which, given the low surface brightness
  ($\sim$23.4~mag arcsec$^{-2}$ in the V band) and large angular
  extent (a core radius of 14~arcmin) of the galaxy (Mateo 1998), is
  probably understandable.}

\tablenotetext{c}{According to \citet{tin97} component B is 5.9~arcsec
  East and 1.6~arcsec S of component A, whose coordinates are given in
  $^d$. In our images we measure similar offset values for component
  B, namely, 5.8~arcsec East and 1.4~arcsec S from component A.}
\end{deluxetable}

\clearpage


\begin{deluxetable}{cccccc}
\tabletypesize{\scriptsize}
\rotate

\tablecaption{Astrometric R-band\tablenotemark{a} frames {\it
    effectively} used in each of our five QSO fields.\label{tbl2}}

\tablewidth{0pt}
\tablehead{
\colhead{QSO ID} & \colhead{Epoch range} & \colhead{\# epochs} & \colhead{\# astrometric frames\tablenotemark{b}} &
\colhead{\# DCR frames} & \colhead{DCR HA range} \\
  & [year]  &  &  &  & [hour]
}
\startdata
QJ0238-3443  & 2000.98 ... 2008.82 & 4 & 09 & 09 &  0.68 ... 2.99  \\
QJ0238-3440  & 2000.98 ... 2008.82 & 4 & 19 & 13 &  0.36 ... 3.86  \\
QJ0239-3420  & 2000.99 ... 2007.87 & 3 & 14 & 12 &  0.55 ... 3.70  \\
QJ0240-3434B\tablenotemark{c} & 2000.61 ... 2007.85 & 3 & 15 & 13 & -0.65 ... -4.03 \\
QJ0240-3438  & 2000.61 ... 2008.83 & 4 & 20 & 10 & -0.74 ... -3.62 \\
\enddata
\tablenotetext{a}{As explained in the text, B-band frames for building
  up CMDs of each field were also acquired.}  \tablenotetext{b}{{\it
    All} of these frames satisfy that $|HA| \le 1.5$~hr, and $FWHM \le
  1.0$~arcsec (as an example, see Table~1 of Paper~I).}
\tablenotetext{c}{Reported in Paper I.}
\end{deluxetable}

\clearpage


\begin{deluxetable}{cccccc}
\tabletypesize{\scriptsize}
\rotate
\tablecaption{Local reference stars and cuts applied in each QSO field.\label{tbl3}}
\tablewidth{0pt}
\tablehead{
\colhead{QSO ID} & \colhead{\# initial LRS stars} & \colhead{\# final LRS stars} & \colhead{$\sigma_{\mu_c}$} &
\colhead{PM error vs. mag} & \colhead{CMD cleansing} \\
       &                      &                    & [$\masy$]       & \# stars purged & \# stars purged
}
\startdata
QJ0238-3443  & 295            & 226                & 0.78            & 3           & 12 \\
QJ0238-3440  & 217            & 175                & 0.60            & 6           &  7 \\
QJ0239-3420  & 337            & 250                & 0.60            & 12          & 11 \\
QJ0240-3434B & 260            & 217                & 0.50            & 11          &  7 \\
QJ0240-3438  & 156            & 123                & 0.60            & 0           &  9 \\
\enddata
\end{deluxetable}

\clearpage


\begin{table}
\begin{center}
\caption{Fornax PM from our five  QSO fields.\label{tbl4}}
\begin{tabular}{ccccc}
\tableline\tableline
QSO ID & $\mu_\alpha \cos \delta$  & $\mu_\delta$ & $\sigma_{fit}$ in RA & $\sigma_{fit}$ in DEC \\
       & [$\masy$]                & [$\masy$]   & [mas]                & [mas]    \\
\tableline
QJ0238-3443  & $0.90 \pm0.15$ & $-0.80 \pm 0.12$ & 1.5 & 1.2 \\
QJ0238-3440  & $0.55 \pm0.13$ & $-0.66 \pm 0.15$ & 1.6 & 2.0 \\
QJ0239-3420  & $0.94 \pm0.17$ & $-0.78 \pm 0.27$ & 1.8 & 2.8 \\
QJ0240-3434B & $0.64 \pm0.08$ & $-0.01 \pm 0.11$ & 0.9 & 1.3 \\
QJ0240-3438  & $0.06 \pm0.21$ & $-0.40 \pm 0.20$ & 2.8 & 2.7 \\
\hline
Unweighted mean & $0.62 \pm0.16$ & $-0.53 \pm 0.15$ & --  & -- \\
\tableline
\tableline
\end{tabular}
\end{center}
\end{table}

\clearpage


\begin{table}
\begin{center}
\caption{Comparison of Fornax PMs derived by \citet{pia07} with our
  results {\it for the same} QSO fields.\label{tbl5}}
\begin{tabular}{ccccc}
\tableline\tableline
QSO ID & $\mu_\alpha \cos \delta$  & $\mu_\delta$ & $\mu_\alpha \cos \delta$  & $\mu_\delta$ \\
       & [$\masy$]                & [$\masy$]   & [$\masy$]                & [$\masy$]    \\
& \multicolumn{2}{c}{$|$---------------------------------$|$} & \multicolumn{2}{c}{$|$---------------------------------$|$} \\
       & \multicolumn{2}{c}{\citet{pia07}} & \multicolumn{2}{c}{This work} \\
\tableline
QJ0238-3443   & $0.45 \pm 0.07$   & $-0.39 \pm 0.06$   & $0.90 \pm0.15$   & $-0.80 \pm 0.12$ \\
QJ0240-3434B  & $0.42 \pm 0.10$   & $-0.48 \pm 0.11$   & $0.64 \pm0.08$   & $-0.01 \pm 0.11$ \\
QJ0240-3438   & $0.54 \pm 0.16$   & $-0.33 \pm 0.16$   & $0.06 \pm0.21$   & $-0.40 \pm 0.20$ \\
\hline
Unweighted mean & $0.47$ & $-0.40$ & $0.53$ & $-0.40$ \\
\tableline
\tableline
\end{tabular}
\end{center}
\end{table}

\clearpage


\begin{deluxetable}{cccccc}
\tabletypesize{\scriptsize}
\rotate
\tablecaption{Published mean Fornax PMs and their uncertainties.\label{tbl6}}
\tablewidth{0pt}
\tablehead{
\colhead{Reference} & \colhead{$\mu_\alpha \cos \delta$} & \colhead{$\mu_\delta$} & \colhead{PM} &
\colhead{PA\tablenotemark{a}} & \colhead{Comments} \\
 & [$\masy$] & [$\masy$] &  [$\masy$] & [deg] &
}
\startdata
\citet{din04}   & $0.59  \pm 0.16$  & $-0.15  \pm 0.16$  & $0.61 \pm 0.16$ & $104 \pm 15$ & Plates, 48 galaxies and 8 QSOs \\
\citet{pia07}   & $0.476 \pm 0.046$ & $-0.360 \pm 0.041$ & $0.597 \pm 0.044$ & $127 \pm 4$ & HST+PC2+STIS, 4 QSO fields \\
This work       & $0.62 \pm 0.16$ & $-0.53 \pm 0.15$ & $0.82 \pm 0.16$ & $125 \pm 11$ & NTT+SuSI2, 5 QSO fields \\
\tableline
\citet{wal08}   & $0.48 \pm 0.15$   & $-0.25 \pm 0.14$   & $0.54 \pm 0.15$ & $118 \pm 15$ &  Radial velocities, ``perspective rotation'' \\
\enddata
\tablenotetext{a}{Position angle measured from North to East.}
\end{deluxetable}

\clearpage


\begin{table}
\begin{center}
\caption{Proper motions in the Galactic coordinate system\tablenotemark{a} and tangential
  velocities\tablenotemark{b} derived from the PMs given in
  Table~\ref{tbl6}.\label{tbl7}}
\begin{tabular}{cccccc}
\tableline\tableline
Reference & $\mu_l$   & $\mu_b$   & $V_l$           & $V_b$ & $V_t$ \\
          & [$\masy$] & [$\masy$] & [km~s$^{-1}$] & [km~s$^{-1}$] & [km~s$^{-1}$] \\
\tableline
\citet{din04} & $0.04 \pm 0.16$ & $0.60 \pm 0.16$ & $31 \pm 112$ & $423 \pm 112$ & $424 \pm 112$ \\  
\citet{pia07} & $0.271 \pm 0.041$ & $0.532 \pm 0.046$ & $189 \pm 31$ & $370 \pm 34$ & $416 \pm 33$  \\  
This work     & $0.41 \pm 0.15$ & $0.70 \pm 0.16$ & $288 \pm 107$ & $490 \pm 113$ & $568 \pm 112$ \\
\tableline
\citet{wal08} & $0.16 \pm 0.14$ & $0.52 \pm 0.15$ & $113 \pm 98$ & $360 \pm 105$ & $377 \pm 104$ \\  
\tableline
\tableline
\end{tabular}

\tablenotetext{a}{To derive Galactic motions from the J2000
  $\mu_\alpha \cos \delta$ and $\mu_\delta$ PMs we have adopted a
  position for the J2000 Galactic Pole of RA=(12:51:26.2754, DEC=
  +27:07:41.705) following \citet{miy93} (see, especially, their
  equation (29)).}  \tablenotetext{b}{To derive the tangential
  velocities $V_l$ and $V_b$, we have a adopted a distance modulus for
  Fornax of $(m-M)_0 = 20.84 \pm 0.15$ from the ``Araucaria distance
  scale project'' (Pietrzy{\'n}ski et al. 2009). The velocity values
  above include a full statistical propagation of PM and distance
  (taken to be $d=147 \pm 10$~kpc) errors. $V_t$ is the total
  Heliocentric tangential velocity, and its error.}
\end{center}
\end{table}

\clearpage


\begin{deluxetable}{ccccccccccc}
\tabletypesize{\scriptsize}
\rotate

\tablecaption{Fornax current Heliocentric\tablenotemark{a} and
  Galactocentric\tablenotemark{b} velocities and specific orbital
  kinematical parameters from values of different
  authors.\label{tbl8}}

\tablewidth{0pt}
\tablehead{
\colhead{Reference} & \colhead{$U_{\mbox{HC}}$} & \colhead{$V_{\mbox{HC}}$} &  \colhead{$W_{\mbox{HC}}$} & \colhead{$V_{r_{\mbox{GC}}}$} 
& \colhead{$V_{l_{\mbox{GC}}}$} 
& \colhead{$V_{b_{\mbox{GC}}}$}  & \colhead{$V_{t_{\mbox{GC}}}$} 
& \colhead{$K/m$\tablenotemark{c}} & \colhead{$L/m$\tablenotemark{d}} & \colhead{$L_z/m$\tablenotemark{d}}   \\
                    &  [km~s$^{-1}$] & [km~s$^{-1}$] & [km~s$^{-1}$] & [km~s$^{-1}$] & [km~s$^{-1}$] & [km~s$^{-1}$] & [km~s$^{-1}$] 
& [$\times 10^{3}$~(km~s$^{-1}$)$^2$] & [$\times 10^{3}$~kpc km~s$^{-1}$ ] & [$\times 10^{3}$~kpc km~s$^{-1}$ ]
}
\startdata
\citet{din04} & $-195 \pm 109$ & $-359 \pm 105$ & $126 \pm 46$ & $-23 \pm 62 $  & $59 \pm 107 $    & $257 \pm 98 $  & $263 \pm  98$ & 35 & 38 & -3.9 \\
\citet{pia07} & $-36 \pm 31$   & $-404 \pm 31$  & $104 \pm 15$ & $-32 \pm 19 $  & $-93 \pm 32 $    & $189 \pm 29 $  & $210 \pm  30$ & 23 & 31 & 6.1 \\
This work     & $-13 \pm 106$  & $-550 \pm 104$ & $154 \pm 47$ & $-33 \pm 62 $  & $-203 \pm 105 $  & $298 \pm 96 $  & $361 \pm  99$ & 66 & 53 & 13.3 \\
\hline
\citet{wal08} & $-95 \pm 97$ & $-355 \pm 96$ & $100 \pm 43$ & $-28 \pm 58 $  & $-16 \pm 97 $  & $186 \pm 89 $  & $ 186 \pm 89$ & 18 & 27 & 1.1 \\
\enddata

\tablenotetext{a}{Heliocentric velocities have been computed using the
  adopted J2000 RA and DEC coordinates for the Fornax center (see
  Table~\ref{tbl7}), which correspond to Galactic coordinates
  $(l,b)=(237.10\degr, -65.65\degr)$. The errors correspond to the
  statitical propagation of all observational errors. The $(U,V,W)$ is
  a right-handed system that points to the Galactic center, Galactic
  rotation and North Galactic Pole respectively.}

\tablenotetext{b}{Galactocentric velocities have been computed
  adopting a solar peculiar veloicity of
  $(u_\odot,v_\odot,w_\odot)=(10.00 \pm 0.36, 5.25 \pm 0.62, 7.17 \pm
  0.38)$~km~s$^{-1}$ from \citet{deh98}, an LSR speed of $220 \pm
  5$~km~s$^{-1}$ and $R_\odot = 8.5 \pm 0.5$~kpc. The errors
  correspond to the statitical propagation of all observational
  errors. While these are commonly used values, we note that recent
  studies seem to suggest a larger V-component of the solar motion
  with respect to the LSR, $\sim$13~km~s$^{-1}$ (e.g., Co{\c s}kuno{\v
    g}lu et al. 2011). However, given the current PM uncertainties,
  this change will have little effect on the calculated motion of
  Fornax. Also, note that our uncertainty for the LSR speed is rather
  optimistic, for example, \citet{bov09} quote $V_{\mbox{LSR}}=244 \pm
  13$~km~s$^{-1}$, while \citet{bru11} argue for $V_{\mbox{LSR}}=239
  \pm 7$~km~s$^{-1}$.}

\tablenotetext{c}{Kinetic energy per unit mass, as seen from a point
  at rest with respect to the Galactic Center.}

\tablenotetext{d}{Total (=$\sqrt{L_x^2 + L_y^2 + L_z^2}$) and
  z-component of the angular momentum per unit mass, with respect to
  the Galactic Center, in our right-handed system.}

\end{deluxetable}

\clearpage


\begin{deluxetable}{cccccccccc}
\tabletypesize{\scriptsize}
\rotate

\tablecaption{Fornax orbital parameters from PM values of different
  authors\tablenotemark{a}.\label{tbl9}}

\tablewidth{0pt}
\tablehead{
\colhead{Reference} & \colhead{Orbital period} & \colhead{Apogalactic distance} & \colhead{Perigalactic distance} & \colhead{$Z_{\mbox{max}}$} & \colhead{$Z_{\mbox{min}}$} &
\colhead{Excentricity} & \colhead{Inclination} & \colhead{$V_Z$} & \colhead{$V_p$} \\ 
 & [Gyr] & [kpc] & [kpc] & [kpc] & [kpc] &  & [deg] & [km~s$^{-1}$] & [km~s$^{-1}$]
}
\startdata
\citet{din04} & $ 8 \pm 4$  & $328 \pm 226$ & $148 \pm 10$ & $223 \pm  84$ & $-238 \pm 108$ & $0.38 \pm 0.21$ & $68 \pm 8$ & $189 \pm  9$ & $265 \pm 51$ \\
\citet{pia07} & $ 6 \pm 1$  & $197 \pm  58$ & $142 \pm 14$ & $163 \pm  35$ & $-166 \pm  37$ & $0.16 \pm 0.09$ & $76 \pm 4$ & $186 \pm  2$ & $221 \pm 17$ \\ 
This work     & $21 \pm 8$  & $956 \pm 376$ & $148 \pm  8$ & $498 \pm 191$ & $-411 \pm 172$ & $0.73 \pm 0.18$ & $41 \pm 8$ & $181 \pm 19$ & $363 \pm 51$ \\ 
\hline
\citet{wal08} & $ 5 \pm 3$  & $164 \pm 109$ & $131 \pm 30$ & $147 \pm  73$ & $-148 \pm  66$ & $0.11 \pm 0.18$ & $83 \pm 7$ & $191 \pm  8$ & $212 \pm 41$ \\
\enddata

\tablenotetext{a}{The errors were computed as the inner 50\%
  interquartile range derived from 1,000 simulations with
  uncertainties drawn from Gaussian distributions of the errors in
  PMs, $V_r$ and distance. The integration period was chosen to be
  large enough (50~Gyr) so that the longest orbital period solution
  completed at least two full orbital cycles.}

\end{deluxetable}


\end{document}